\documentclass[twocolumn]{aastex62}
\usepackage{amsmath}
\graphicspath{{./}{figures/}}

\shorttitle{Interstellar Meteors are Outliers in Material Strength}
\shortauthors{Siraj \& Loeb}

\begin{document}

\title{Interstellar Meteors are Outliers in Material Strength}

\email{amir.siraj@cfa.harvard.edu, aloeb@cfa.harvard.edu}

\author{Amir Siraj}
\affil{Department of Astronomy, Harvard University, 60 Garden Street, Cambridge, MA 02138, USA}

\author{Abraham Loeb}
\affiliation{Department of Astronomy, Harvard University, 60 Garden Street, Cambridge, MA 02138, USA}




\begin{abstract}
The first interstellar meteor larger than dust was detected by US government sensors in 2014, identified as an interstellar object candidate in 2019, and confirmed by the Department of Defense in 2022. Here, we describe an additional interstellar object candidate in the CNEOS fireball catalog, and compare the implied material strength of the two objects, referred to here as IM1 and IM2, respectively. IM1 and IM2 are ranked 1 and 3 in terms of material strength out of all 273 fireballs in the CNEOS catalog. Fitting a log-normal distribution to material strengths of objects in the CNEOS catalog, IM1 and IM2 are outliers at the levels of $3.5 \sigma$ and $2.6 \sigma$, respectively. The random sampling and Gaussian probabilities, respectively, of picking two objects with such high material strength from the CNEOS catalog, are $\sim 10^{-4}$ and $\sim 10^{-6}$. If IM2 is confirmed, this implies that interstellar meteors come from a population with material strength characteristically higher than meteors originating from within the solar system. Additionally, we find that if the two objects are representative of a background population on random trajectories, their combined detections imply that $\sim 40\%$ of all refractory elements are locked in meter-scale interstellar objects. Such a high abundance seemingly defies a planetary system origin. 
\end{abstract}



\section{Introduction}

CNEOS\footnote{\url{https://cneos.jpl.nasa.gov/}} 2014-01-08, detected by U.S. Department of Defense (DoD) sensors through the light that it emitted as it burned up in the Earth's atmosphere off of the coast of Papua New Guinea in 2014, was determined to be an interstellar object in 2019 \citep{2019arXiv190407224S}, a conclusion that was confirmed by independent analysis conducted by the DoD in 2022 \citep{shaw_2022}, although the numerical uncertainties were not provided to the scientific community. The object, which we refer to here as IM1, predated the interstellar object `Oumuamua by 3.8 years, and the interstellar object Borisov by 5.6 years. The measured peak flare apparent in the light curve of IM1 at an altitude of $18.7 \mathrm{\; km}$ implies ambient ram pressure of $\sim 194$ MPa when the meteor disintegrated \citep{2022RNAAS...6...81S}. This level of material strength is $\gtrsim 20$ times higher than stony meteorites and $\gtrsim 2$ times larger than iron meteorites. IM1 was also dynamically unusual -- its speed relative to the Local Standard of Rest (LSR) is shared by less than 5\% of all stars. 

In this \textit{Letter}, we describe an interstellar meteor candidate from the CNEOS catalog mentioned in Table 8 of \cite{2022AJ....164...76P} and in Section 5 of \cite{2022ApJ...939...53S}, which we refer to as IM2. We then explore the statistical likelihood that interstellar meteors reflect the same distribution of material strength as non-interstellar meteors. 

\section{Interstellar Meteor Candidate}

The Python code implemented here used the open-source N-body integrator software \texttt{REBOUND}\footnote{https://rebound.readthedocs.io/en/latest/} to trace the motion of the meteor under the gravitational influence of the Solar System \citep{2012A&A...537A.128R}.

We initialize the simulation with the Sun, the eight planets, and the meteor, with geocentric velocity vector $(vx_{obs}, vy_{obs}, vz_{obs}) = (-15.3, 25.8, -20.8) \; \mathrm{km\;s^{-1}}$, located at $40.5^{\circ}$ N $18.0^{\circ}$ W, at an altitude of 23.0 km, at the time of impact, $t_i =$ 2017-03-09 04:16:37 UTC, as reported in the CNEOS catalog. We then use the IAS15 adaptive time-step integrator to trace the meteor's motion back in time \citep{2015MNRAS.446.1424R}. This does not account for air drag, which would lead to an even higher impact speed, and therefore heliocentric speed, given the encounter geometry. The slowdown of IM1 due to air drag was estimated in earlier work \citep{2022RNAAS...6...81S}.

There are no substantial gravitational interactions between the meteor and any planet other than Earth for any trajectory within the reported errors. Based on the geocentric impact speed reported by CNEOS, $v_{obs} = 36.5\; \mathrm{km\;s^{-1}}$, the heliocentric impact speed was $\sim 50 \; \mathrm{km\;s^{-1}}$. We find that the meteor was unbound with an asymptotic speed of $v_{\infty} \sim 25.9\; \mathrm{km\;s^{-1}}$ outside of the solar system. 

We find the heliocentric orbital elements of the meteor at impact to be the following: semi-major axis, $a = -1.1 \;$ AU, eccentricity, $e = 1.6$, inclination $i = 26 \;$ degrees, longitude of the ascending node, $\Omega = -12$ degrees, argument of periapsis, $\omega = 241 \;$ degrees, and true anomaly, $f = 300 \;$ degrees. It was $v_{LSR} \sim 40\; \mathrm{km\;s^{-1}}$ away from the velocity of the LSR \citep{2010MNRAS.403.1829S}.

Given the explosion energy of $\sim 4 \times 10^{19} \; \mathrm{\; ergs}$ and the atmospheric impact speed of $\sim 36.5 \mathrm{\; km \; s^{-1}}$, we adopt equivalence between the pre-explosion kinetic energy and the energy in the explosion, finding that the object's mass was $\sim 6.3 \times 10^6 \mathrm{\; g}$. A comparison between the properties of IM1 and IM2 is included in Table \ref{tab:table}.

We note that the fact that both IM1 and IM2 have low orbital inclinations ($i \lesssim 30^{\circ}$) is puzzling, since interstellar objects are expected to have a uniform distribution in $\cos{i}$. Specifically, the random likelihood of two orbital inclinations of $i \lesssim 30^{\circ}$ drawn from a uniform distribution in $\cos{i}$ is, $(1 - \cos{(30^{\circ})})^2 \sim 2 \%$. However, 2I/Borisov had an inclination of $44^{\circ}$, and the likelihood of drawing three inclinations of $i \lesssim 45^{\circ}$ is $(1 - \cos{(45^{\circ})})^2 \sim 3 \%$, so there may be an inclination bias in the source of a certain class of interstellar objects. 1I/`Oumuamua had an orbital inclination of $i = 122^{\circ}$, more indicative of a background population uniform distribution.

\begin{figure}
 \centering
\includegraphics[width=1\linewidth]{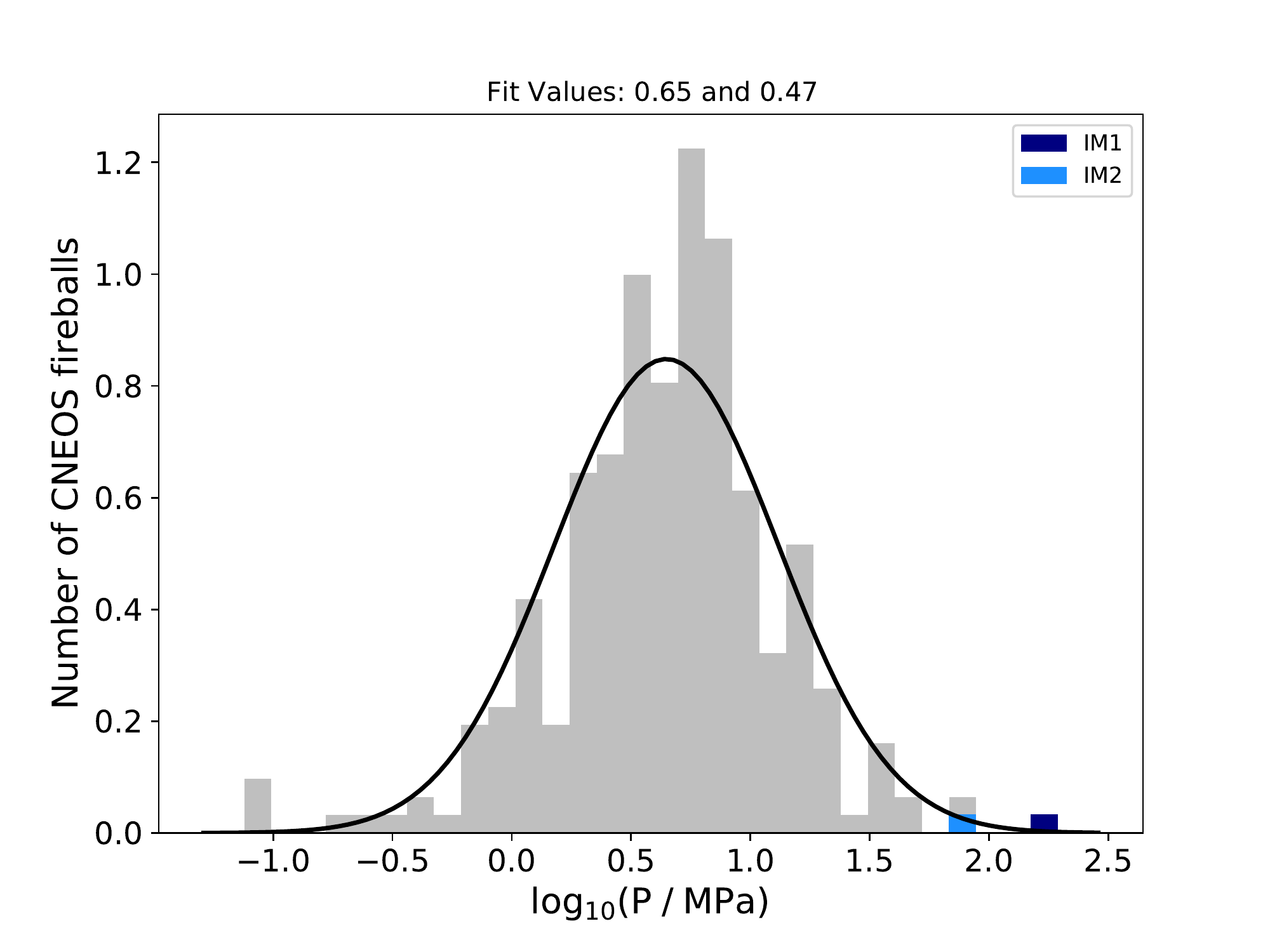}
\caption{Histogram of ram pressure at breakup, $\rho v^2$, for all fireballs in the CNEOS catalog with altitudes and velocities reported at the times of peak brightness. Navy blue and bright blue correspond to IM1 and IM2, respectively, which are ranked 1 and 3 in terms of material strength out of all 273 fireballs. Black line indicates the best fit to a Gaussian with mean and standard deviation matching the data, $\mu = 0.47$ and $\sigma = 0.65$.}
\label{fig:histogram}
\end{figure}

\begin{figure}
 \centering
\includegraphics[width=1\linewidth]{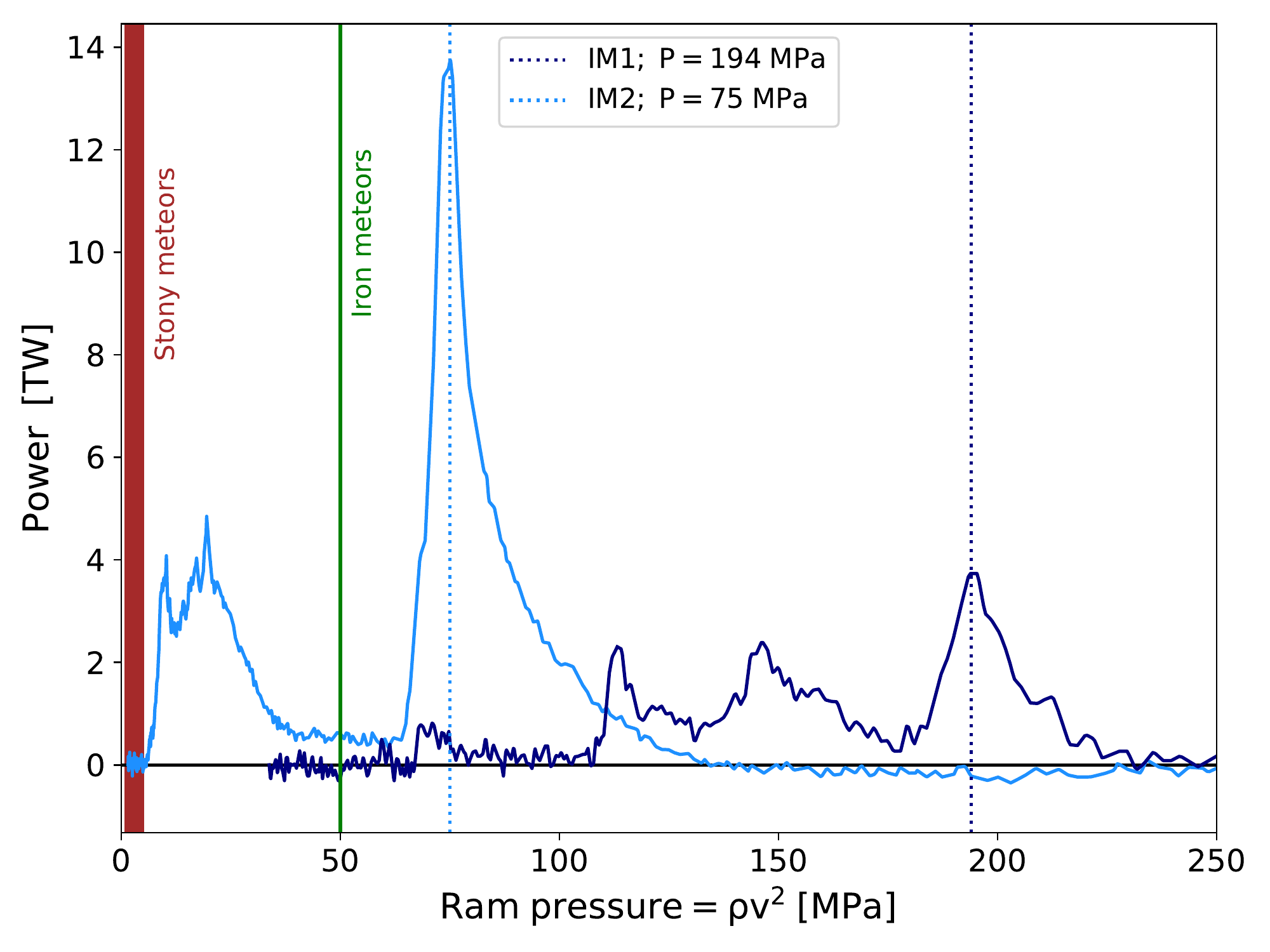}
\caption{Total power released in the IM1 and IM2 fireballs (navy blue and bright blue, respectively) as a function of ram pressure, $\rho v^2$. Peak brightness is reached at 194 MPa and 75 MPa, respectively, for the two fireballs. Typical stony and iron meteorite yield strengths, $1 - 5 \mathrm{\; MPa}$ and $50 \mathrm{\; MPa}$ respectively, are indicated for convenience of comparison. Note that 1 TW = $10^{19} \; \mathrm{erg \; s^{-1}}$ and 1 MPa = $10^7 \mathrm{\; dyne \; cm^{-2}}$. The IM2 light curve is calibrated based on the total energy released (taking account of a normalization error in the published light curve). There is an early flare in the IM2 light curve indicating some amount of lower-strength material, in addition to the clearly central flare corresponding to a metallic composition.}
\label{fig:ram_pressure}
\end{figure}

\section{Material Strength Comparison}

\begin{deluxetable*}{cccccccc}
\tabletypesize{\footnotesize}
\tablecolumns{8}
\tablewidth{0pt}
\tablecaption{Comparison between IM1 and IM2. \label{tab:table}}
\tablehead{
\colhead{Designation} & \colhead{$t_i$} & \colhead{$v_{obs} \mathrm{\; (km/s)}$} &\colhead{$v_{\infty}\mathrm{\; (km/s)}$} & \colhead{$v_{LSR} \; \mathrm{\; (km/s)}$}  & \colhead{$Y_{i} \mathrm{\; (MPa)}$} & \colhead{$m \mathrm{\; (g)}$} & \colhead{$n \mathrm{\; (AU^{-3})}$}}
\startdata
IM1 & 2014-01-08 17:05:34 & 44.8 & 42.1 & 60 & 194 & $4.6 \times 10^5$ & $1.8 \times 10^6$\\ 
IM2 & 2017-03-09 04:16:37 & 36.5 & 25.9 & 40 & 75 & $6.3 \times 10^6$ & $2.7 \times 10^6$
\enddata
\vspace{-0.8cm}
\end{deluxetable*}

As a meteor travels through the atmosphere, it experiences friction due to air. Dynamical pressure is $\rho v^2$, and the dynamical pressure corresponding to the peak power in the meteor light curve describes the material strength of the meteor, since crossing a certain level of ram pressure causes the object to deform and break apart \citep{2011MPS...46.1525P, 2006MNRAS.372..655T, 2007MNRAS.375..415T}.

Based on estimates for comets, carbonaceous, stony, and iron meteorites \citep{1993Natur.361...40C, 1993Natur.365..733S, 1995Icar..116..131S, 2001JMatS..36.1579P}, \cite{2005M&PS...40..817C} established an empirical strength-density relation for impactor density $\rho_i$ in the range $1 - 8 \mathrm{\; g \; cm^{-3}}$. The upper end of this range gives a yield strength of $Y_i \sim 50 \mathrm{\; MPa}$, corresponding to the strongest known class of meteorites, iron \citep{2001JMatS..36.1579P}. Iron meteorites are rare in the solar system, making up only $\sim 5\%$ of modern falls \citep{2006mess.book..869Z}.

We computed the ram pressure at breakup for all 273 fireballs in the CNEOS catalog. Interestingly, IM1 and IM2 display the first and third highest material strengths, respectively, amongst all of the fireballs. Figure \ref{fig:histogram} is a histogram showing all of the fireballs in the catalog, and highlighting IM1 and IM2. Figure \ref{fig:ram_pressure} shows the light curves for IM1 and IM2, with the peak ram pressures highlighted. 

The probability of randomly drawing 2 of the top 3 material strengths, out of all 273 fireballs, is $\sim (3 / 273)^2 \sim 10^{-4}$. Therefore, if IM2 is confirmed to be an interstellar meteor, simple random drawing dictates that there would be a $\sim 99.99\%$ chance that interstellar meteors come from a population with material strength characteristically higher than meteors originating from within the solar system, a notion suggested by \cite{2022AJ....164...76P}.

The material strength data follows a log-normal distribution, shown in Figure \ref{fig:histogram} with mean $\mu = 0.65$ and $\sigma = 0.47$. MPa. As a result, IM1 and IM2 represent $3.5 \sigma$ and $2.6 \sigma$ deviations from the mean, respectively. These deviations correspond to $2.4 \times 10^{-4}$ and $4.5 \times 10^{-3}$ single-tailed probabilities, respectively. Combining these independent events, we find a $\sim 10^{-6}$ probability of getting the material strengths of IM1 and IM2 by random chance. This Gaussian perspective implies a $\sim 99.9999\%$ chance that interstellar meteors are characteristically stronger than meteors from within the solar system. 

We note that the ram pressure could be overestimated by the fact that DoD satellites only detected the brightest sections of their luminous paths. For instance, the US government sensor data on the Chelyabinsk event gives a dynamic strength twice or three times as large as that measured in the recovered meteorites \citep{2022AJ....164...76P}.

\newpage

\section{Implications for Local Mass Budget}

\begin{figure}
  \centering
  \includegraphics[width=\linewidth]{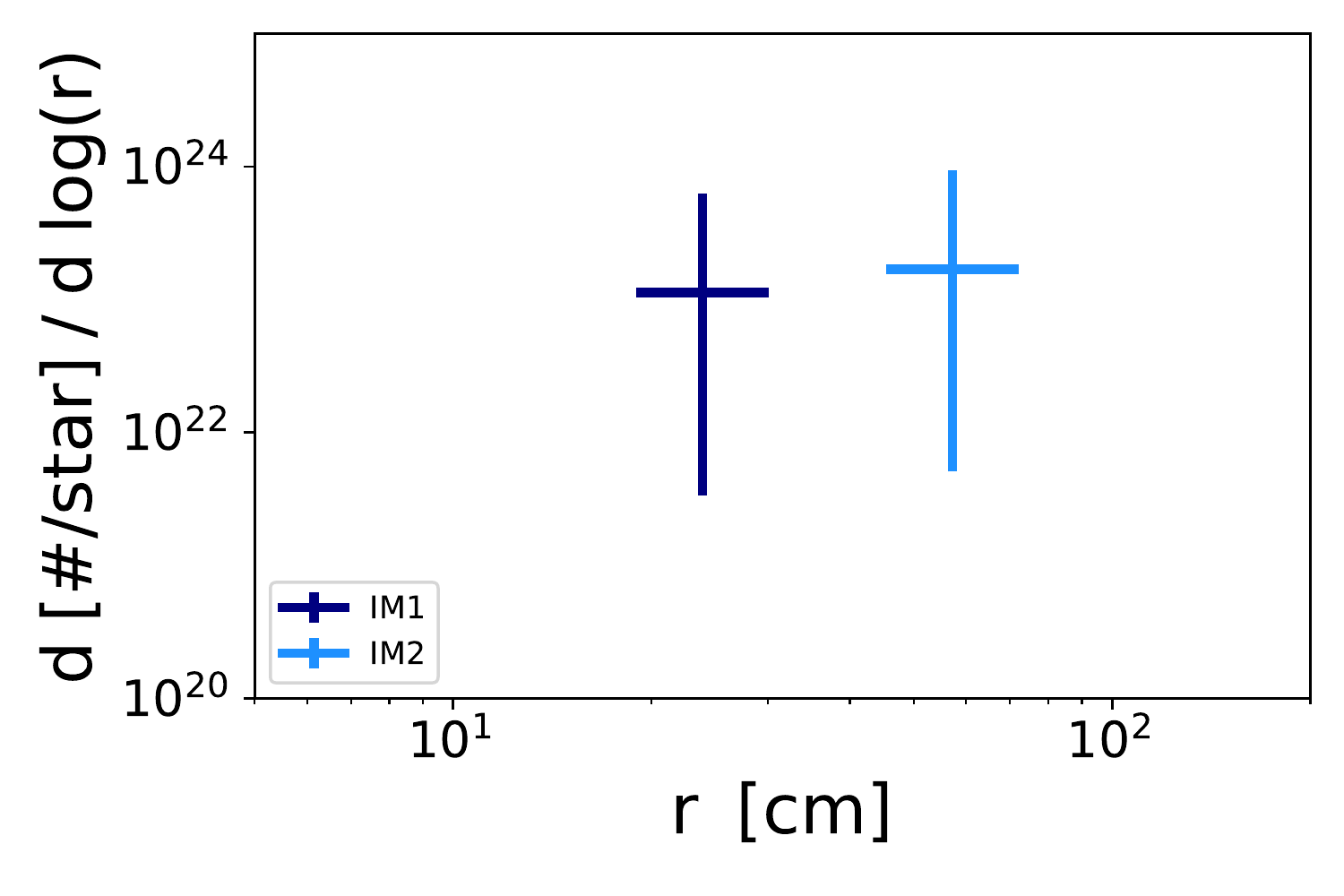}
    \caption{IM1 and IM2 in size-abundance parameter space, expressed as number per star per differential unit of log size. The vertical error bars correspond to the 95\% Poisson uncertainties, while the horizontal error bars correspond to a factor of two in mass, in each direction.}
    \label{fig:sizedist}
\end{figure}

For a background population on random trajectories drawn from an isotropic distribution in the LSR, the number density implied by the detection of an interstellar meteor is,
\begin{equation}
n \simeq \frac{\Gamma}{v_{\infty} \pi R_{\oplus}^2 [1 + (v_{esc} / v_{\infty})^2]} \; \; ,
\end{equation}
where $\Gamma$ is the implied rate, $v_{\infty}$ is the speed outside of the solar system,\footnote{Without knowing the velocity dispersion of the population, our best estimate is taking the velocity of the object relative to the Sun as a ``typical relative speed," including the velocity of the Sun relative to the LSR and the velocity dispersion of the population in the LSR. We only have a sample of two  objects which are not sufficient for a better statistical analysis to find the velocity dispersion of the population.} $R_{\oplus}$ is the radius of the Earth, and $v_{esc} = \sqrt{2 G M_{\odot} / d_{\oplus}}$ is the escape speed from the Earth's position, where $M_{\odot}$ is the mass of the Sun and $d_{\oplus}$ distance between the Earth and the Sun. Gravitational focusing is accounted for by the term in the denominator, $[1 + (v_{esc} / v_{\infty})^2]$. We adopt $\Gamma \sim 0.1 \mathrm{\; yr^{-1}}$ for both IM1 and IM2 \citep{2019arXiv190407224S}, and speeds outside of the solar system of $v_{\infty} \sim 42 \mathrm{\; km \; s^{-1}}$ and $v_{\infty} \sim 26 \mathrm{\; km \; s^{-1}}$, respectively. We find that the number density implied the detections of IM1 and IM2 are $n_{IM1} \sim 1.8 \times 10^6 \mathrm{\; AU^{-3}}$ and $n_{IM2} \sim 2.7 \times 10^6 \mathrm{\; AU^{-3}}$, as shown in Figure \ref{fig:sizedist}.

Given the respective masses of $\sim 4.6 \times 10^5 \mathrm{g}$ and $\sim 6.3 \times 10^6 \mathrm{g}$, we find that the detections of IM1 and IM2 imply, respectively, ambient local abundances of $\sim 1.2 \mathrm{\; M_{\oplus} \; pc^{-3}}$ and $\sim 25 \mathrm{\; M_{\oplus} \; pc^{-3}}$ of similar objects. 

The local stellar mass density is $\sim 0.04 \mathrm{\; M_{\odot} \; pc^{-3}}$ \citep{2017MNRAS.470.1360B}. The local density of the interstellar medium is $1.2 \mathrm{\; cm^{-3}}$ \citep{2015ApJ...814...13M}, implying $\sim 0.03 \mathrm{\; M_{\odot} \; pc^{-3}}$. All refractory elements (metals and silicates) sum to a total mass fraction of $\sim 0.3 \%$ at solar metallicity, implying that the local budget of metals and silicates in stars and dust is $\sim 70 \mathrm{\; M_{\oplus} \; pc^{-3}}$. We conservatively assume that IM1 and IM2 are composed of refractory elements, even though their material strengths imply that they were primarily metallic in composition. If IM2 is indeed an interstellar object, the detections of IM1 and IM2 together imply that $\sim 40\%$ of all refractory elements locked from stars and the ISM are locked in meter-scale interstellar objects. Refractory elements are also locked in objects bound to the Sun, with the largest theorized reservoir being the Oort cloud \citep{1990ApJ...359..506S}.

\section{Discussion}

If interstellar meteors are formed in planetary systems, the natural limit to the scale of mass ejected is the total budget of the minimum mass solar nebula model (MMSN), which is of order $\sim 1 \%$ of stellar mass \citep{1977MNRAS.180...57W, 1981PThPS..70...35H, 2007ApJ...671..878D, 2009ApJ...698..606C}. The result reached here indicates that if IM2 is confirmed an interstellar object, the detections of IM1 and IM2 combined imply that $\sim 2/3$ of the mass budget in stars is necessary to provide the refractory elements to produce a population of interstellar meteors that would make the detections of IM1 and IM2 likely. This result thereby provides a new constraint on planetary system formation, since it requires nearly two orders of magnitude more mass than the MMSN \citep{1977MNRAS.180...57W, 1981PThPS..70...35H, 2007ApJ...671..878D, 2009ApJ...698..606C}. Note that the mass budget discussed exceeds that of objects `Oumuamua-sized and larger, which itself is an unsolved puzzle \citep{2021arXiv211105516S, 2021arXiv211015213L}. The extraordinary mass budget required to produce interstellar meteors seemingly defies planetary system origins, and suggests some other highly efficient route for creating meter-scale objects made of refractory elements. Interestingly, there is a paucity of refractory elements observed in the gas phase in the interstellar medium \citep{1996ARAA..34..279S, 2005AA...429..297M, 2009ApJ...694.1335D}, an observation which could potentially reflect refractory elements being locked in interstellar objects. Supernovas have been observed to produce iron-rich ``bullets", which could be a possible origin of IM1 and IM2 \citep{1994astro.ph..5071L, 1995Natur.377..315S, 1995Natur.373..590S, 2002ApJ...574..155W, 2006NewAR..50..521T, 2009LPI....40.1999P, 2013MNRAS.430.2864M, 2015MNRAS.453..166T, 2021ApJ...921..113S}.

\section*{Acknowledgements}
This work was supported in part by a grant from the Breakthrough Prize Foundation and by research funds from the Galileo Project at Harvard University. 





\bibliography{bib}{}
\bibliographystyle{aasjournal}



\end{document}